\documentclass{article}

\usepackage{arxiv}

\usepackage[utf8]{inputenc} 
\usepackage[T1]{fontenc}    
\usepackage{hyperref}       
\usepackage{url}            
\usepackage{booktabs}       
\usepackage{amsfonts}       
\usepackage{nicefrac}       
\usepackage{microtype}      
\usepackage{lipsum}		
\usepackage{graphicx}
\usepackage{natbib}
\usepackage{float}
\usepackage[font=small,labelfont=bf]{caption}

\title{Dreamento: an open-source dream engineering toolbox for sleep eeg wearables}

\usepackage{authblk}
\author[1,*]{Mahdad Jafarzadeh Esfahani}
\author[1,2]{Amir Hossein Daraie}
\author[1]{Paul Zerr}
\author[1,3]{Frederik D. Weber}
\author[1]{Martin Dresler}
\affil[1]{Donders Institute for Brain, Behaviour and Cognition, Radboudumc, Nijmegen, The Netherlands}
\affil[2]{Department of Biomedical Engineering, Johns Hopkins University School of Medicine, Baltimore,
Maryland, USA}
\affil[3]{Department of Sleep and Cognition, Netherlands Institute for Neuroscience, an institute of the Royal
Netherlands Academy of Arts and Sciences, Amsterdam, The Netherlands}
\affil[*]{Corresponding author}

\date{}

\renewcommand{\headeright}{}
\renewcommand{\undertitle}{}
\renewcommand{\shorttitle}{\textit{Dreamento:} an open-source dream engineering toolbox for sleep EEG wearables }

\begin{document}
\maketitle

\begin{abstract}
We introduce Dreamento (Dream engineering toolbox), an open-source Python package for dream
engineering using sleep electroencephalography (EEG) wearables. Dreamento main functions are (1) real-time recording, monitoring, analysis, and sensory stimulation, and (2) offline post-processing of the resulting data, both in a graphical user interface (GUI). In real-time, Dreamento is capable of (1) data recording, visualization, and navigation, (2) power-spectrum analysis, (3) automatic sleep scoring, (4) sensory stimulation (visual, auditory, tactile), (5) establishing text-to-speech communication, and (6) managing annotations of automatic and manual events. The offline functions aid in post-processing the acquired data with features to reformat the wearable data and integrate it with non-wearable recorded modalities such as electromyography (EMG). While Dreamento was primarily developed for (lucid) dreaming studies, its applications can be extended to other areas of sleep research such as closed-loop auditory stimulation and targeted memory reactivation.
\end{abstract}


\section{Introduction}

The gold standard of measuring human sleep is polysomnography (PSG) which consists of electroencephalography (EEG), electromyography (EMG), and electrooculography (EOG) as the primary physiological signals. The American Academy of sleep medicine (AASM) also recommends recording additional modalities such as electrocardiography (ECG), blood oxygen saturation level, and body position when studying sleep \citep{iber2007aasm}. Such full PSG setups provide great data quality but are accompanied by several constraints, including limitations of artificial lab environments and the time, effort, and cost invested by researchers attaching electrodes and sensors. Furthermore, standard sleep scoring based on PSG recordings is not only resource demanding but also subject to considerable variability in inter-rater agreement \citep{rosenberg2013american,danker2009interrater}.
\par
Based on advancements in miniature electronics, various wearable systems such as smart watches, smart rings, and EEG headbands have recently been launched in the consumer technology market. Given the prominence of scalp EEG in studying sleep, EEG headbands in particular have received substantial attention from sleep researchers. While wearable systems overcome some limitations of PSG, they have their own restrictions. Headbands typically utilize an EEG montage that is different from AASM standards which makes human scoring more challenging. Furthermore, most EEG headbands process the data directly on the device, i.e., onboard processing, preventing more accurate but resource-demanding real-time computations. Automatic sleep scoring and sleep modulation (e.g., to apply sensory stimuli during a specific stage of sleep) using such onboard computations are in most cases not very reliable due to limited onboard computational resources. Less affected by these constraints are wearable systems with cloud computing features or the ability to communicate with a computer in real-time, enabling extensive processing through the use of dedicated computer resources. The ZMax (Hypnodyne Corp., Sofia, Bulgaria) sleep wearable is an example of an EEG headband that employs a transmission control protocol/internet protocol (TCP/IP) to transmit data to a computer in real-time. This gives considerable freedom to developers and researchers to design software for a variety of purposes and can therefore make the performance of wearable systems more reliable.
\par 
To serve as a reliable alternative to full PSG setups, supplementary analysis tools are needed so that the output of wearables can make up for their shortcomings compared to full PSG setups. Several open-source sleep analysis toolboxes are available, e.g., tools to visualize and analyze offline sleep data such as SleepTrip (RRID: SCR\textunderscore017318, \url{https://github.com/Frederik-D-Weber/sleeptrip}, Sleep \citep{combrisson2017sleep}, Visbrain \citep{combrisson2019visbrain}, YASA \citep{vallat2021open}, and various open-source automatic sleep scoring algorithms (e.g. \citep{perslev2021u,supratak2017deepsleepnet, supratak2020tinysleepnet, vallat2021open}). The current literature, however, lacks open-source tools to monitor, analyze, and modulate sleep in \textit{real-time}. To this end we developed an open-source dream engineering toolbox with unique features.
\par
Exploiting the existing features of the ZMax headband, we developed an open-source, Python-based toolbox with graphical user interface (GUI) for dream engineering, dubbed Dreamento (Dream engineering toolbox, \url{https://github.com/dreamento/dreamento}) that works both online in real-time and offline. By introducing Dreamento, we intend to facilitate sleep and dream research by providing a standard tool for performing experiments with minimal sensing systems in real-life environments, and with large sample sizes. \textit{Real-time} Dreamento is developed to record, monitor, analyze, and modulate sleep in real-time, whereas \textit{offline} Dreamento provides tools for post-processing of the resulting data. The most notable features of real-time Dreamento are (1) data recording, visualization, and navigation, (2) power-spectrum analysis (periodogram and time-frequency representation), (3) automatic sleep scoring, (4) sensory stimulation (visual, auditory, and tactile with the desired properties of the stimulus), (5) text-to-speech communication, and (6) saving annotations of the automatic and manual events. Offline Dreamento enables post-processing of the acquired data (either recorded using real-time Dreamento or the official software of ZMax Hypnodyne headband, i.e., HDRecorder) by employing similar features to real-time Dreamento such as (1) data visualization, navigation, and scalability, (2) power-spectrum analysis, (3) and automatic sleep scoring using more robust algorithms (given that performance speed while post-processing is less critical than in real-time analysis). Additionally, (4) offline Dreamento is capable of integrating the simultaneously acquired data (e.g., EMG recording through another device) with the ZMax data recorded through HDRecorder and real-time Dreamento and (5) exporting the raw and processed results. Importantly, Dreamento is not restricted to the ZMax platform, but will be transferable to other sleep EEG wearables with real-time control functionality. 


\section{Materials and Methods}
\subsection{Programming language and dependencies}
Dreamento was implemented in Python 3 which has stable open-source packages as a basis to build on and upon. A list of Dreamento's dependencies on external libraries can be found on the Dreamento Github page (\url{https://github.com/dreamento/dreamento/blob/main/dreamento.yml} and \url{https://github.com/dreamento/dreamento/blob/main/offlinedreamento.yml} for the real-time and offline Dreamento, respectively). The toolbox can be installed via Conda (\url{https://conda.io}), an open-source environment manager to create a virtual environment based on the required dependencies (instructions can be found on the Github page). Our package is developed and tested on a Windows 10, 64 bit computer with 16 GB RAM but is also compatible with macOS, and Linux. Although we developed Dreamento in Python, due to a large number of MATLAB (Mathworks, Natick, Massachusetts, USA) users, we have also given the option to export all raw and processed data from Dreamento into MATLAB for further analysis. 
\subsection{Documentation}
Our toolbox is delivered with a detailed step-by-step documentation, from how to install and use the toolbox, to a detailed description of its programming classes, methods, and functions useful for developers. In this way, we facilitate the contribution of software developers and researchers to extend the applications of Dreamento. The documentation can be found at \url{https://dreamento.github.io/docs/}.
\subsection{Hypnodyne software suite}
The producer of the ZMax headband, Hypnodyne, provides a software suite including HDFormat, HDScorer, HDServer, and HDRecorder (which can be freely downloaded from the official website of the company, \url{https://hypnodynecorp.com}). For the purpose of real-time data representation, the ZMax headband can be wirelessly connected to the computer through a USB dongle. The HDServer initiates the TCP/IP server and HDRecorder operates as the main client of the server, capable of displaying and recording various signals such as the two EEG channels, triaxial acceleration, photoplethysmography (PPG), body temperature, nasal airflow (additional sensor required), ambient noise, ambient light, and battery consumption. 
\par
Some functionalities which are desirable for (lucid) dream engineering studies are not supported by HDRecorder and have thus been implemented in Dreamento. (1) In real-time recording, the time and amplitude axes should be adjustable (e.g., to set up the desired amplitude limit dynamically while detecting different microstructural features of sleep). (2) Since some sleep events (e.g., rapid eye movement; REM) are of short duration, the user should be able to navigate back in the data display (at least within the current 30-second epoch) so as to confirm the event. (3) Information regarding the sensory stimulation such as stimulus type, properties, and the exact time of presentation should be automatically stored. (4) For real-time sleep scoring, the experimenter should be able to flexibly implement different autoscoring algorithms. (5) It should be possible to include manual annotations when a remarkable event happens. (6) Additional signal qualities, e.g., power-spectrum and time-frequency representation (TFR) should be provided as complementary information for real-time scoring and analysis of sleep.
\subsection{Program structure}
As shown in table \ref{tab1}, Dreamento comprises different programming classes, namely \textit{ZmaxSocket}, \textit{ZmaxDataID}, \textit{ZmaxHeadband}, \textit{Window}, \textit{RecordThread}, \textit{DreamentoScorer}, and \textit{OfflineDreamento}. We defined the configuration of connection to the TCP/IP server (e.g., host IP address and the port number) in \textit{ZmaxSocket}. In addition, this class is responsible for establishing two-way communication between the client and the server, i.e., data chunk transmission from the server to the client and sending commands/messages such as stimulation properties from the client to the server. To enhance the code’s readability, \textit{ZmaxDataID} enumerates data that can be collected (e.g., EEG channels and triaxial accelerometer) with a specific identity number. 
\textit{ZmaxHeadband} class is capable of either (1) decoding the sensor readings such as the EEG or acceloremeter outputs and converting them into interpretable values or (2) encoding communicative messages such as the stimulation commands and sending them to the server. The initialization of buffer sizes for each data channel is also incorporated in this class. 
\par

\begin{table}
	\caption{The main functions of different programming classes used in the development of Dreamento.\label{tab1}}
	\centering
	\begin{tabular}{ll}
		\multicolumn{2}{c}{}               \\
        \toprule
        \textbf{Programming class} & \textbf{Function}  \\
        \midrule
        ZmaxSocket		& Establishing a connection to the TCP/IP server		\\
        ZmaxDataID		& Enumerating each data signal with a specific identity number		\\
        ZmaxHeadband & Encoding commands to the headband and decoding the received data \\
        Window & Defining the main functionalities of the GUI window \\
        RecordThread & Transmitting data to the main window when ready for plotting and analysis\\
        DreamentoScorer & Offline automatic sleep scoring \\
        OfflineDreamento & Post-processing of the recorded data \\
        \bottomrule
	\end{tabular}
	\label{tab1}
\end{table}


All variables related to the data recording (e.g., which signals to record), monitoring (e.g., time and amplitude scales of signals), analysis (e.g., activation of real-time autoscoring), and stimulation (e.g., stimulus properties) are specified in the \textit{Window} class. This class also determines the functions associated with all the GUI buttons, from a primary button that activates the connection to the server, to buttons for triggering stimulation commands (see section \ref{section_GUI}). We designed \textit{RecordThread} as a thread to fetch the data in real-time and send it to the \textit{Window} as well as to maintain the accurate timing of the processes. This was done so that while displaying data in real-time, the user can simultaneously employ other features of the toolbox, such as stimulation or annotation assignments. The thread keeps track of the received number of samples from the server (as a measure of passed time) and once an epoch of 30 seconds (equivalent to 7680 samples based on 256 Hz sampling rate) is over, activates the corresponding flag to indicate that the buffer to analyze the data is ready. The data chunk will be subsequently sent to the \textit{Window} where the relevant analysis such as the autoscoring and TFR are performed.
\par
Post-processing functions, from loading the data to the generation of results are integrated in \textit{OfflineDreamento} class. This class also uses the \textit{DreamentoScorer} which is implemented for offline autoscoring (see details in section \ref{section_autoscoring}). 

\subsection{Automatic sleep scoring}
\label{section_autoscoring}
In this research, we have used a validated autoscoring algorithm i.e., TinySleepNet \citep{supratak2020tinysleepnet} for \textit{real-time} autoscoring in Dreamento. This algorithm is based on a convolutional neural network (CNN) and a long short-term memory (LSTM) stack, and provides sleep stage predictions after completion of each 30-second epoch. This algorithm was chosen for real-time Dreamento due to its acceptable prediction speed of 13 ms on average (which plays an essential role in real-time analysis).
\par
As for offline Dreamento, given that processing time of the algorithm is less critical and accuracy is more important, we propose \textit{DreamentoScorer}. This algorithm first extracts various time, frequency, time-frequency, linear, and non-linear features of each epoch of the EEG signals (see Dreamento Github page for a complete list of features). Then, in order to take time dependence into account (as in conventional manual scoring), for the prediction of each epoch's sleep stage, it concatenates features not only from the currently viewed epoch but also from a few previous epochs (number of epoch is user adjustable). Afterwards, a feature selection algorithm, i.e., Boruta \citep{kursa2010feature} keeps the relevant features and removes the redundant ones. Eventually, a LightGBM \citep{ke2017lightgbm} model is used to predict the sleep stage.
\par
To train the current versions of the real-time and offline autoscoring algorithms, we used a dataset consisting of 69 nocturnal sleep datasets collected with ZMax and a standard PSG system simultaneously (publication in preparation). In order to provide the algorithms with accurately labeled data (sleep stages) for training, the PSG data were scored by a human expert, after which the scoring was aligned with the corresponding ZMax recording.
\unskip
\subsection{Experimenter pipeline}
\begin{figure}[H]
\includegraphics[width= \linewidth]{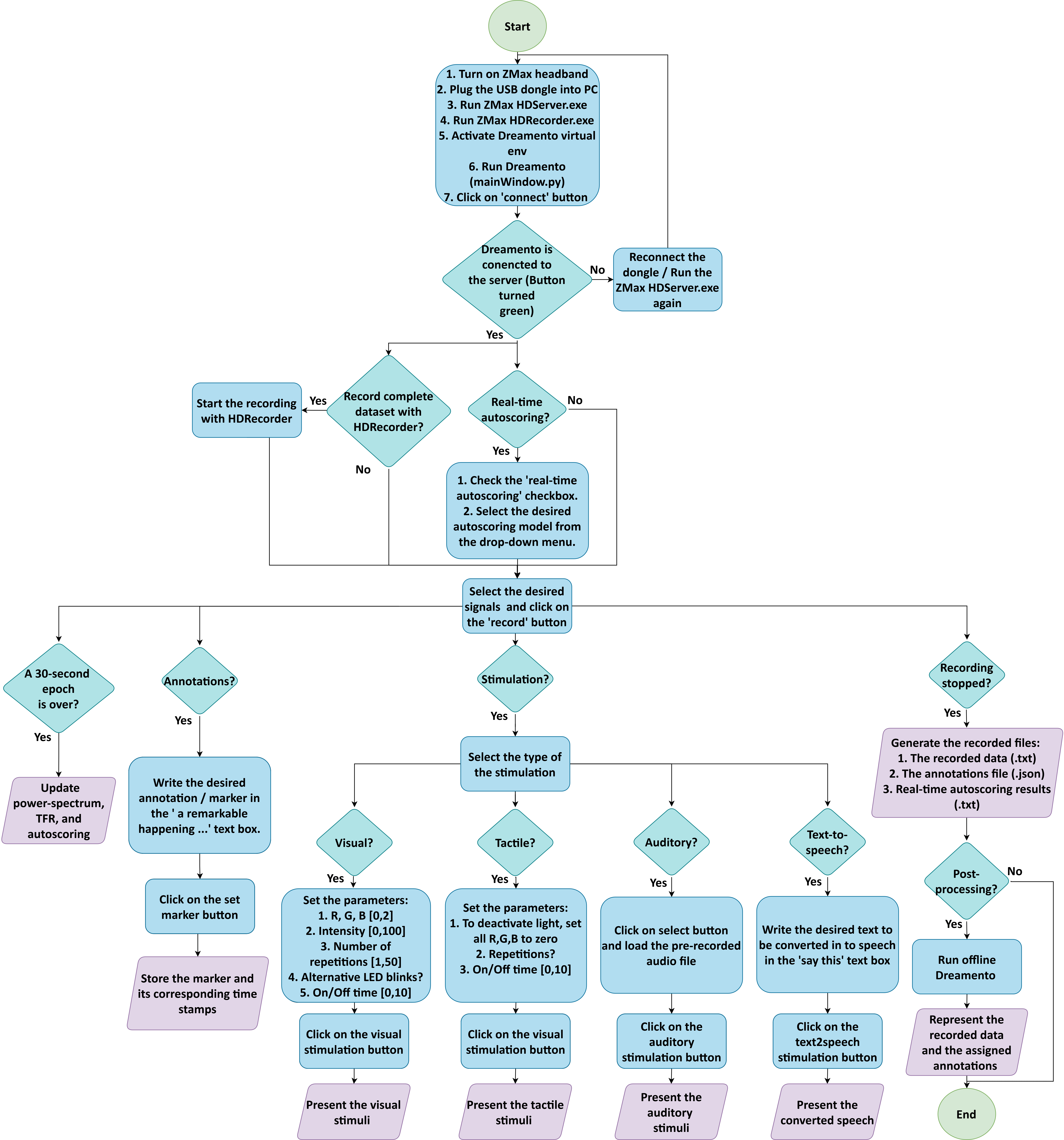}
\caption{Real-time Dreamento experimenter pipeline to record, monitor, analyze, and stimulate sleep. LED: light emitting diode, TFR: time-frequency representation, R: red, G: green, B: blue. \label{fig1}}
\end{figure}

Figure \ref{fig1} shows the experimenter pipeline of real-time Dreamento. Once the connection to the server is established, the recording can be started by clicking the ‘record’ button. After the start of the recording, the user is able to set the desired annotations and apply sensory stimulation at any time. Furthermore, Dreamento updates the analytical outcomes such as autoscoring, power-spectral density, and TFR whenever an epoch of 30 s is over. Once the recording is stopped, the output files comprising the recorded data, annotation files, and real-time scoring results will be generated. 

\par
\begin{figure}[H]
\begin{center}
\includegraphics[width=14 cm]{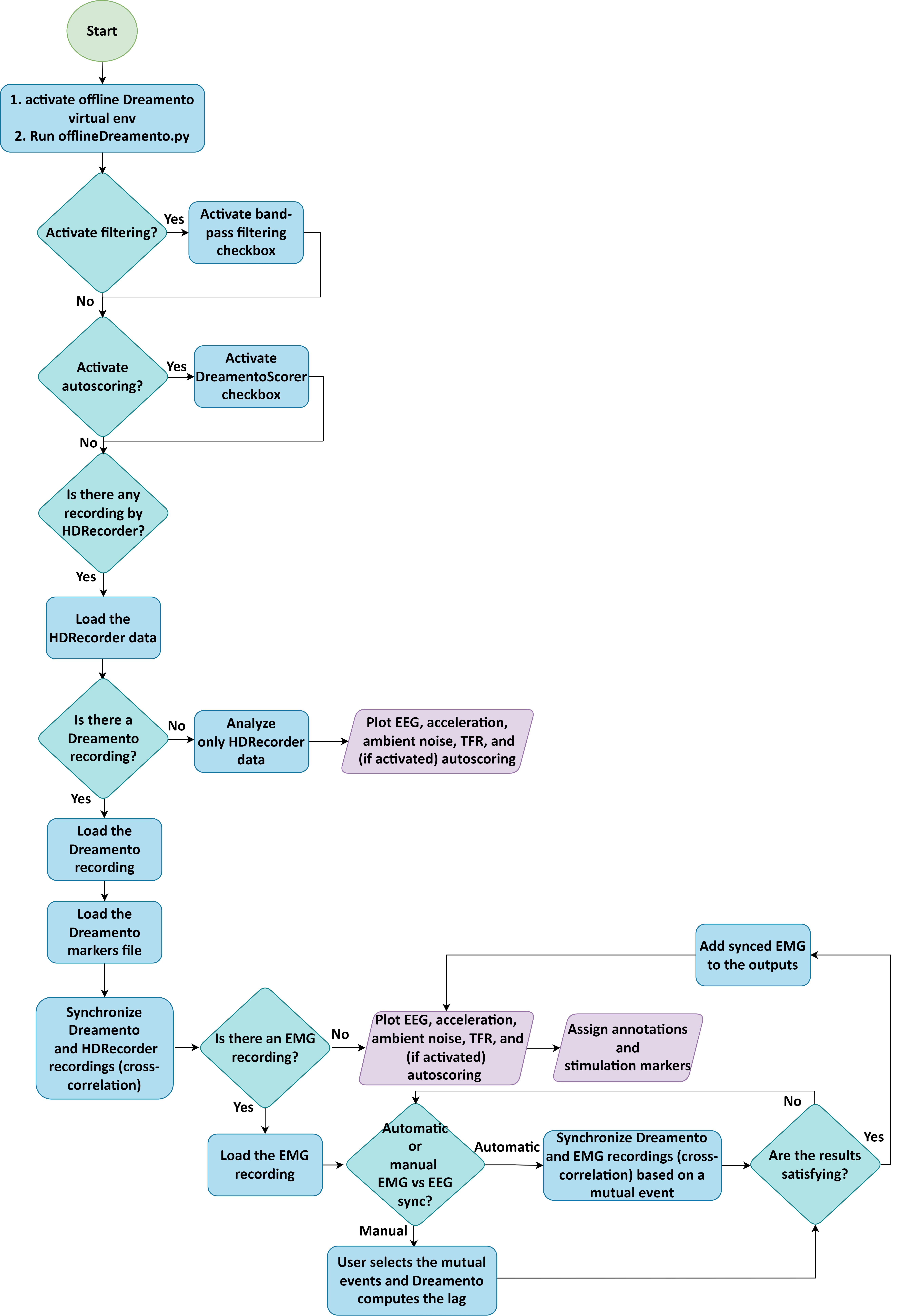}
\end{center}
\caption{Offline Dreamento experimenter pipeline. EEG: electroencephalography, EMG: electromyography, TFR: time-frequency representation.
\label{fig_offline_Dreamento_pipeline}}
\end{figure}

Offline Dreamento provides the user with the opportunity to post-process the data collected by ZMax: (1) either using HDRecorder or real-time Dreamento, or (2) using both HDRecorder and real-time Dreamento (simultaneously), and (3) simultaneous recordings with HDRecorder, real-time Dreamento, and other measurement modalities such as an EMG system (see the pipeline in Figure \ref{fig_offline_Dreamento_pipeline}). Depending on the types of  input data loaded by the user in Dreamento, the software synchronizes the recordings (see section \ref{section_sync} for details), performs the desired analysis and subsequently represents the output (e.g., autoscoring, TFR, annotations, and stimulation markers). 
 
\newpage

\newpage

\newpage

\subsection{Graphical user interface (GUI)}
\label{section_GUI}
\begin{figure}[h]
\begin{center}
\includegraphics[width=14 cm]{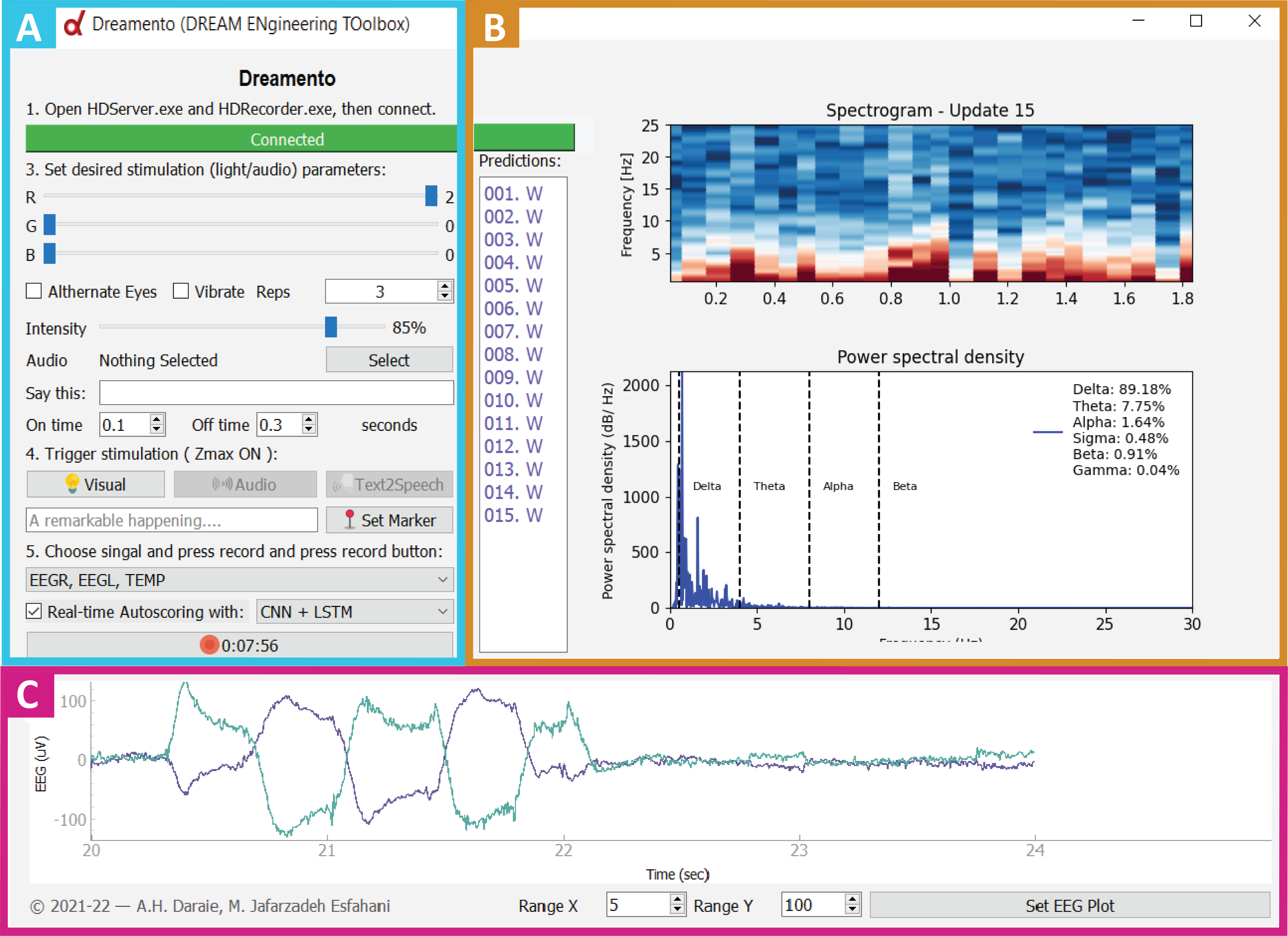}
\end{center}
\caption{GUI of real-time Dreamento. \textbf{(A)} the recording and stimulation panel: this panel is used for connecting Dreamento to the server, selecting the desired signals to record (e.g., EEG and temperature), activating the autoscoring, and both configuring and presenting the desired sensory stimulation. \textbf{(B)} analysis panel: where the autoscoring, power spectral analysis and TFR update after every 30-second epoch completion (the TFR keeps the data from approximately last 2 minutes for a better representation of the recent sleep stage transitions), and \textbf{(C)} real-time data representation panel: the real-time EEG data are depicted in this panel. The time (x-axis) and amplitude (y-axis) ranges can be changed while recording data.\label{fig2}}
\end{figure}

\begin{figure}[h]
\includegraphics[width= \linewidth]{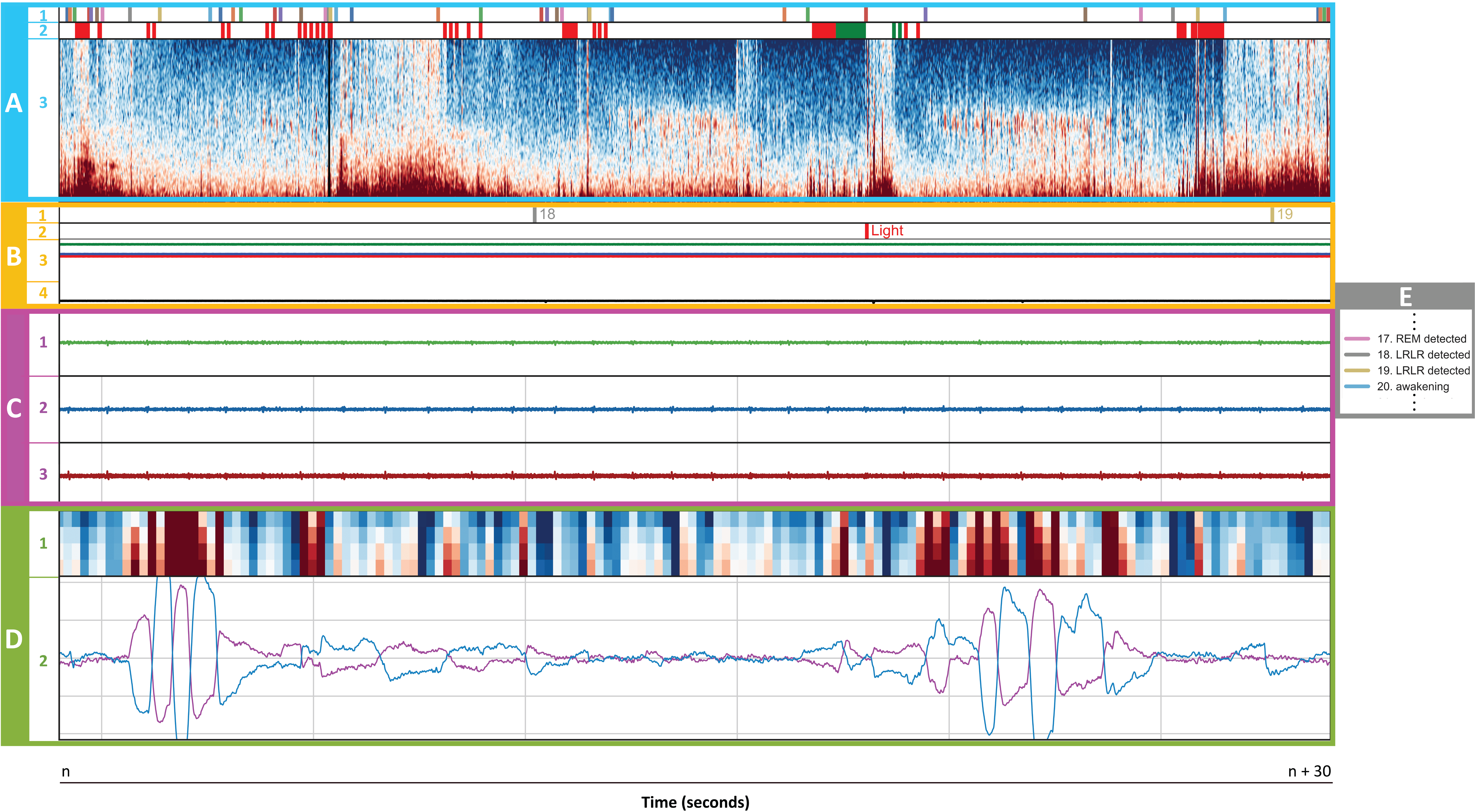}
\caption{GUI of offline Dreamento (units, values, and labels on the vertical axes are removed for a clearer representation). Panel \textbf{(A)} depicts the overall recording representation of (1) manual annotations, (2) automatic annotations, i.e., stimulus presentation (red: visual, green: tactile, blue: auditory stimulus), and (3) the corresponding TFR of the complete recording. The black line in the TFR shows the currently selected epoch. Panels \textbf{(B)} to \textbf{(D)} refer to the current epoch, as indicated at the bottom, from the second \textit{n} to \textit{n + 30}. Panel \textbf{(B)} shows the distribution of the (1) manual annotations, (2) stimulus presentation, (3) triaxial acceleration indicated by red, green, and blue (for the x, y, and z axes), and (4) ambient noise (the flat black line represents no ambient noise/sound). Panel \textbf{(C)} shows three EMG channels recorded by a BrainAmp ExG system simultaneously with ZMax integrated into offline Dreamento for post-processing. The EEG representation of the selected epoch together with its corresponding TFR is shown in panel \textbf{(D)}. The list of manual annotations is shown in panel \textbf{(E)}.\label{fig_offline_Dreamento}}
\end{figure}

The user interfaces of the real-time and offline Dreamento are illustrated in Figures \ref{fig2} and \ref{fig_offline_Dreamento}, respectively. In real-time Dreamento, the recording and stimulation panel (Figure \ref{fig2}, panel A) contains the relevant parameters to (1) start a recording (e.g., the type of the data to collect, for instance EEG and temperature) as well as (2) configure and apply sensory stimulation (visual, auditory, tactile), and (3) establish text-to-speech communication. For example, to present a visual stimulus, it is possible to set up the desired color of the light-emitting diodes (LEDs), select the required intensity (brightness) of light, choose the number of LED blinks, determine whether the two LEDs of the headband should blink simultaneously or alternatively, and set the on/off timing of the LEDs. As shown in Figure \ref{fig2}, panel B, the real-time analysis panel consists of the TFR, periodogram, and the autoscoring predictions. The autoscoring and periodogram keep the outcome of the last 30-second epoch only, whereas the TFR maintains the output from the last four epochs (approximately two minutes). This helps the experimenter to have an estimate of the sleep stage transitions over the past few epochs. While recording, the software depicts real-time EEG signals with adjustable scales for time and amplitude axes as shown in Figure \ref{fig2}, panel C.

By stopping the data recording, the software generates three output files, namely, the actual recorded raw data (.txt), annotations (.json), and real-time scoring (.txt) results. Given the inconsistency of the sampling rates during a wireless communication (the fluctuations from the actual 256 Hz sampling rate of the headband), for an accurate estimation of actual time, the number of transmitted samples per second is stored in addition to the recorded data. The annotation file stores all manually set annotations (e.g., remarkable events) and stimulation markers, properties, and their corresponding time stamps. 
\par

Offline Dreamento integrates different sources of data/information and represents them in a GUI (Figure \ref{fig_offline_Dreamento}). The top three rows of the window (Figure \ref{fig_offline_Dreamento} – panel A) are assigned to the annotations, stimulation markers, and TFR of the \textit{complete} recording. Thus, with a glance at the first three rows of the display, the user gains an overview of the annotation distributions, stimulation types and timing (shown with red, blue, and green for visual, auditory, and tactile stimulations, respectively), as well as an estimate of sleep stage transitions using the TFR. All the rest of the rows (Figure \ref{fig_offline_Dreamento}, panels B to D) correspond to the single epoch determined by the black vertical line shown in the overall TFR (Figure \ref{fig_offline_Dreamento}, A-3). These rows represent the annotations (Figure \ref{fig_offline_Dreamento}, B-1), stimulation markers (Figure \ref{fig_offline_Dreamento}, B-2), triaxial acceleration (Figure \ref{fig_offline_Dreamento}, B-3), ambient noise (Figure \ref{fig_offline_Dreamento}, B-4), three EMG channels recorded simultaneously using another device (Figure \ref{fig_offline_Dreamento}, panel C), TFR (Figure \ref{fig_offline_Dreamento}, D-1), and the two EEG channels (Figure \ref{fig_offline_Dreamento}, D-2) of the selected 30-second epoch. For a better mapping between the annotation descriptions and their corresponding time stamps, a specific color and a number are assigned to each (Figure \ref{fig_offline_Dreamento}, panel E).

\subsection{Data synchronization}
\label{section_sync}

\begin{figure}[H]
\includegraphics[width= \linewidth]{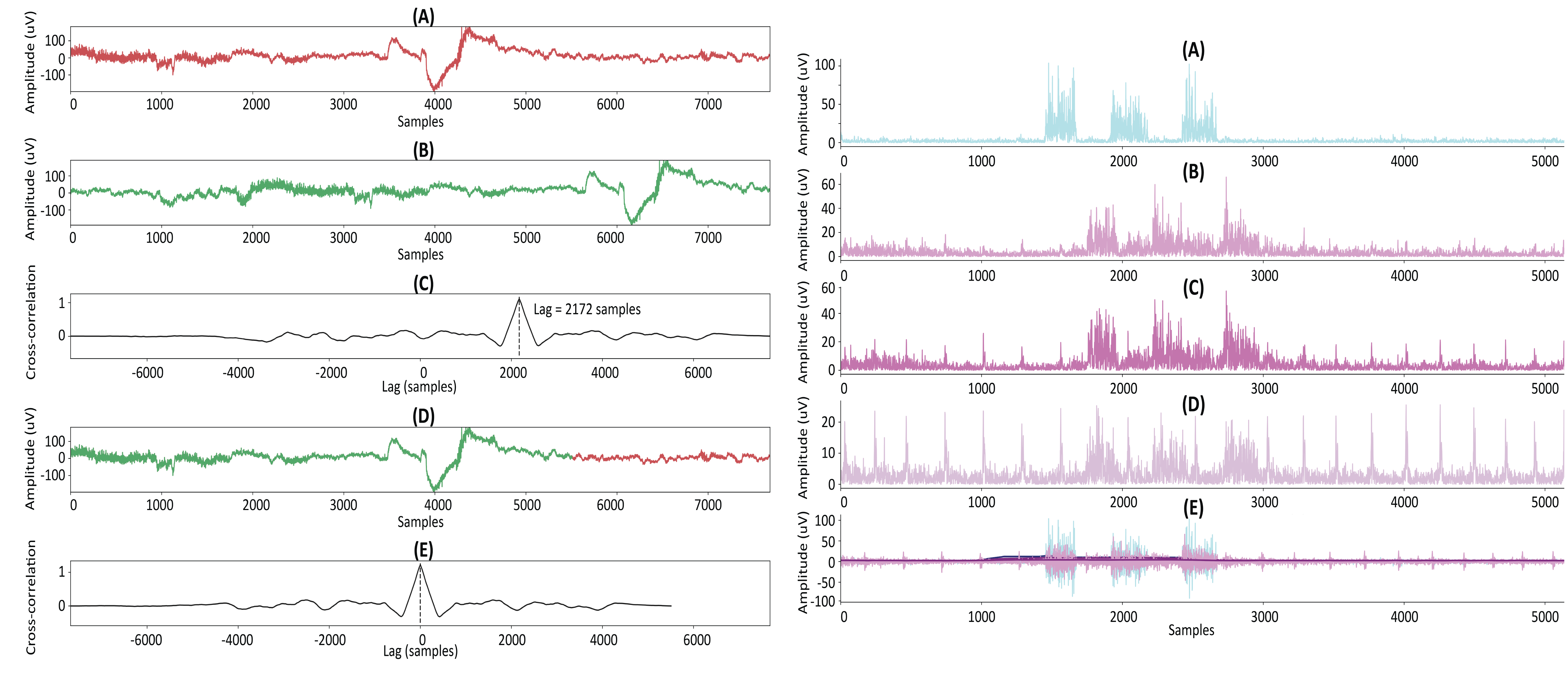}
\caption{Synchronization procedure during post-processing. \textbf{Left panel:} Real-time Dreamento and HDRecorder signals synchronization. First, a sample epoch of 30 seconds (i.e., 7680 samples) from the recorded \textbf{(A)} Dreamento (red signal) and \textbf{(B)} HDRecorder (green signal) was chosen. \textbf{(C)} Cross-correlating the signals in \textbf{(A)} and \textbf{(B)} resulted in a peak in the cross-correlation function which corresponded to the lag (2172 samples in this case) between the signals. \textbf{(D)} Dreamento and HDRecorder signals plotted on top of each other after the synchronization based on the cross-correlation results and \textbf{(E)} proof of no lag (i.e., full alignment) between the signals at the end of the synchronization process. \textbf{Right panel:} EEG and EMG synchronization. \textbf{(A)} EEG signal during performing the teeth clenching task, \textbf{(B)} EMG channel 1, \textbf{(C)} EMG channel 2, and \textbf{(D)} EMG channel 1 - channel 2 during performing the teeth clenching task, \textbf{(E)} the synchronized signals based on the proposed method. \label{fig_sync}}
\end{figure}

Dreamento is capable of recording a subset of the data that the ZMax headband can provide (e.g., EEG and acceleration). To record all data available on the device (i.e., EEG, acceleration, ambient noise, light, temperature, and heart rate), the experimenter can simultaneously record with HDRecorder (see Figure \ref{fig1}). As there will always be a time difference between the initialization of the two programs, we developed a post-processing synchronization algorithm to align Dreamento and HDRecorder recordings (Figure \ref{fig_sync} - left panel). The synchronization process starts by loading EEG data recorded by both programs. Next, Dreamento selects a portion of the recorded data (e.g., from 100 - 130 seconds as in default settings) and computes a cross-correlation analysis, under the assumption that the difference between the start of recordings is less than 30 seconds. Otherwise, the user can adjust the number of seconds based on an overview of the recorded data. Dreamento identifies the lag corresponding to the maximum amplitude of the cross-correlation function and shifts the HDRecorder recording to align it with Dreamento's recording. 

\par

In addition to the data provided by the ZMax headband, Dreamento enables the user to synchronize and integrate different physiological signals such as EMG that have been recorded simultaneously with real-time Dreamento through other devices, e.g., BrainAmp  (Brainproducts GmbH, Gilching, Germany) systems (see Figure \ref{fig_sync}, right panel). The synchronization process between EEG and EMG starts by searching for a user-defined annotation, e.g., teeth clenching, which corresponds to an event that presents in a relatively similar way in both EEG and EMG. Next, the toolbox depicts the actual EEG and EMG signals in the vicinity of the synchronization event (Figure \ref{fig_sync}, right panel A to D) without any alignment attempt and asks the user whether the signals need further alignment. If the user selects to proceed with the synchronization, Dreamento provides two options of either automatic or manual alignment. The former is similar to the correlation-based analysis for aligning HDRecorder and Dreamento signals with some differences. Given that the EMG signal includes high-frequency activities (typically above 10 Hz), Dreamento applies a band-pass filter on the EEG and EMG signals (e.g., between 10 – 100 Hz), determines the absolute values of signals, applies a moving average filter to smoothen the data, and finally, a correlation-based analysis computes the lag between the signals. Nonetheless, if the events in either the EEG and EMG signals are not entirely clear (e.g., when the baseline EMG activity during the emergence of the events is relatively high), the automatic alignment may need some manual adjustments. Therefore, the user is also provided with a manual alignment option. The manual alignment option asks the user to click on the starting points of the same events presented in both EEG and EMG and then compensates for the lag between the recordings based on the selected time points.

\par

\section{Discussion}
Research into lucid dreaming typically faces one of the following limitations: (1) being constrained to self-reports and lacking physiological measurements for verification, (2) small sample sizes (low power), and (3) lack of generalizability by being constrained to a laboratory environment \citep{jafarzadeh2022wearable}. The complexity of standard PSG systems and the need for sleep research personnel in particular makes studying lucid dreaming in the laboratory very resource intensive, which can explain the small sample sizes typically encountered in the literature. Given the availability of minimal sensing systems such as wearable EEG headbands and thanks to recent developments in artificial intelligence, these systems could now extensively be used to resolve some of the controversial findings in the field by conducting large scale (and even at-home) studies. To date no software package exists that allows researchers to easily interface with EEG wearables and analyze sleep data in real-time. To this end, we developed Dreamento in order to simplify sleep and (lucid) dreaming research. Dreamento is an all-in-one package for recording, monitoring, stimulating, and analyzing sleep data in real-time in addition to post-processing the collected data.
\par
Recently, we have started a multi-center lucid dreaming induction study using Dreamento (pre-registered on the Open Science Framework, \url{https://osf.io/u5m3z}) to assess the applicability of a combination of cognitive training and sensory stimulation approach to induce lucid dreams \citep{jafarzadeh2022wearable}. In addition to evaluating a novel approach to induce lucid dreams, the performance of Dreamento will be validated in this multi-center study, which comprises the largest sample size for a lucid dreaming induction study including physiological measures to date (overall 60 planned participants). This in-lab study will then be followed by a multi-center, fully automatic lucid dreaming induction study which participants conduct at home using features of Dreamento.
\par
Other areas of research that can be explored using the sensory stimulation features of Dreamento are dream communication studies that aim at establishing a medium between the (lucid) dreamer and the outside world \citep{konkoly2021real} or even non-REM sleep modulation studies that employ techniques such as targeted memory reactivation (TMR), e.g., \citep{rudoy2009strengthening,rasch2007odor} or closed-loop auditory stimulation (CLAS) \citep{ngo2013auditory}. Of note, the potential clinical applications of these non-invasive stimulation techniques already magnified the importance of developing open hardware (e.g., ZMax headband) and software (e.g., Dreamento) to translate these methods into a naturalistic home setting \citep{jafarzadehesfahani_farboud_ngo_schneider_weber_talamini_dresler_2022}. 

Real-time sleep data analysis faces certain challenges. Real-time algorithms need to be both accurate and fast enough to keep up with the incoming data stream. In our study, real-time analysis comprised autoscoring, time-frequency representations, and periodogram updates after every epoch of 30 seconds. This means that every 30 seconds, the toolbox does not receive new input from the server for a short period of time (e.g., 15 ms), during which it runs the relevant analysis on the last epoch. While the program is busy with the real-time analysis and thus closes the port for new data entry, the data transmitted from the server remains in the queue to enter the program as soon as the analysis is completed. Therefore, if we simply let the queued data (which will be accumulated over time) enter the software, the program will no longer work in actual 'real-time'. To solve this limitation, Dreamento ignores the small portion of the data that remains in queue during the processing time (e.g., 15 ms) of real-time analysis and thus always remains in sync with the real-time data received from the server, regardless of the duration of recording. Nevertheless, we recommend parallel data recording using HDRecorder in order to also store the small missing data chunks in Dreamento's recording.
\par
Despite the fact that a variety of automatic sleep scoring algorithms have been presented to date, it is still not possible to nominate one as superior to the rest. This is mainly due to the fact that each algorithm is sensitive to a specific EEG montage and measurement system. To tackle this limitation, recently a few autoscoring algorithms such as \citep{perslev2021u, vallat2021open} have been validated on large datasets comprising a combination of EEG data collected from different locations, with various EEG systems and based on different montages. Although these algorithms were shown to be effective on various EEG systems and montages, their performance on wearable EEG systems (which typically employ non-conventional EEG montages, e.g., F7/F8 – Fpz as in ZMax) remains unknown. This demonstrates the importance of developing an autoscoring algorithm specifically for wearable EEG systems. In this study, we introduced \textit{DreamentoScorer} as an open-source tool for autoscoring of the ZMax headband (autoscoring validation publication in preparation) and retrained a validated model, i.e., TinySleepNet based on ZMax recordings. Nevertheless, the generalizability of these algorithms can still be enhanced in future work by considering larger training sets.

\section{Conclusions}
In this paper, we present a new open-source dream engineering toolbox, Dreamento (\url{https://github.com/dreamento/dreamento}) that provides a range of tools for recording, monitoring, stimulating, and analyzing sleep data in real-time. The toolbox also includes post-processing functions that allow users to analyze the collected data. Minor changes in Dreamento code can make it compatible with other EEG systems (particularly wearables) that are able to transmit real-time data to Python. This means that sleep researchers interested in real-time signal analyses such as spectrogram or power-spectrum analysis may adapt Dreamento to their use case. Dreamento's user-friendly interface allows experimenters to conduct sleep experiments in an interactive and intuitive manner. Of note, the application of Dreamento is not limited to (lucid) dreaming studies as it can assist with sensory stimulation in other research areas such as in closed-loop auditory stimulation or targeted memory reactivation during non-REM sleep.

\newpage

\bibliographystyle{apalike-ejor}
\bibliography{references}  

\begin{thebibliography}{}

\bibitem[Combrisson et~al., 2017]{combrisson2017sleep}
Combrisson, E., Vallat, R., Eichenlaub, J.-B., O'Reilly, C., Lajnef, T.,
  Guillot, A., Ruby, P.~M., \& Jerbi, K. (2017).
\newblock Sleep: an open-source python software for visualization, analysis,
  and staging of sleep data.
\newblock {\em Frontiers in neuroinformatics}, 60.
\newblock \url{https://doi.org/10.3389/fninf.2017.00060}

\bibitem[Combrisson et~al., 2019]{combrisson2019visbrain}
Combrisson, E., Vallat, R., O'Reilly, C., Jas, M., Pascarella, A., Saive,
  A.-l., Thiery, T., Meunier, D., Altukhov, D., Lajnef, T., et~al. (2019).
\newblock Visbrain: a multi-purpose gpu-accelerated open-source suite for
  multimodal brain data visualization.
\newblock {\em Frontiers in Neuroinformatics}, 13, 14.
\newblock \url{https://doi.org/10.3389/fninf.2019.00014}

\bibitem[Danker-hopfe et~al., 2009]{danker2009interrater}
Danker-hopfe, H., Anderer, P., Zeitlhofer, J., Boeck, M., Dorn, H., Gruber, G.,
  Heller, E., Loretz, E., Moser, D., Parapatics, S., et~al. (2009).
\newblock Interrater reliability for sleep scoring according to the
  rechtschaffen \& kales and the new aasm standard.
\newblock {\em Journal of sleep research}, 18(1), 74--84.
\newblock \url{https://doi.org/10.1111/j.1365-2869.2008.00700.x}

\bibitem[Esfahani et~al., 2022a]{jafarzadeh2022wearable}
Esfahani, M.~J., Carr, M., Salvesen, L., Picard-Deland, C., Demšar, E.,
  Daraie, A., Keuren, Y., McKenna, M.~C., Konkoly, K., Weber, F.~D., Schoch,
  S., Bernardi, G., \& Dresler, M. (2022a).
\newblock Lucid dream induction with sleep eeg wearables.
\newblock {\em OSF}.
\newblock \url{https://doi.org/10.17605/OSF.IO/U5M3Z}

\bibitem[Esfahani et~al.,
  2022b]{jafarzadehesfahani_farboud_ngo_schneider_weber_talamini_dresler_2022}
Esfahani, M.~J., Farboud, S., Ngo, H.-V.~V., Schneider, J., Weber, F.~D.,
  Talamini, L., \& Dresler, M. (2022b).
\newblock Closed-loop auditory stimulation of sleep slow oscillations: basic
  principles and best practices.
\newblock {\em PsyArXiv}.
\newblock \url{https://doi.org/10.31234/osf.io/7xtds}

\bibitem[Iber, 2007]{iber2007aasm}
Iber, C. (2007).
\newblock The aasm manual for the scoring of sleep and associated events:
  Rules.
\newblock {\em Terminology and Technical Specification}.

\bibitem[Ke et~al., 2017]{ke2017lightgbm}
Ke, G., Meng, Q., Finley, T., Wang, T., Chen, W., Ma, W., Ye, Q., \& Liu, T.-Y.
  (2017).
\newblock Lightgbm: A highly efficient gradient boosting decision tree.
\newblock {\em Advances in neural information processing systems}, 30.

\bibitem[Konkoly et~al., 2021]{konkoly2021real}
Konkoly, K.~R., Appel, K., Chabani, E., Mangiaruga, A., Gott, J., Mallett, R.,
  Caughran, B., Witkowski, S., Whitmore, N.~W., Mazurek, C.~Y., et~al. (2021).
\newblock Real-time dialogue between experimenters and dreamers during rem
  sleep.
\newblock {\em Current Biology}, 31(7), 1417--1427.
\newblock \url{https://doi.org/10.1016/j.cub.2021.01.026}

\bibitem[Kursa \& Rudnicki, 2010]{kursa2010feature}
Kursa, M.~B. \& Rudnicki, W.~R. (2010).
\newblock Feature selection with the boruta package.
\newblock {\em Journal of statistical software}, 36, 1--13.
\newblock \url{https://doi.org/10.18637/jss.v036.i11}

\bibitem[Ngo et~al., 2013]{ngo2013auditory}
Ngo, H.-V.~V., Martinetz, T., Born, J., \& M{\"o}lle, M. (2013).
\newblock Auditory closed-loop stimulation of the sleep slow oscillation
  enhances memory.
\newblock {\em Neuron}, 78(3), 545--553.
\newblock \url{https://doi.org/10.1016/j.neuron.2013.03.006}

\bibitem[Perslev et~al., 2021]{perslev2021u}
Perslev, M., Darkner, S., Kempfner, L., Nikolic, M., Jennum, P.~J., \& Igel, C.
  (2021).
\newblock U-sleep: resilient high-frequency sleep staging.
\newblock {\em NPJ digital medicine}, 4(1), 1--12.
\newblock \url{https://doi.org/10.1038/s41746-021-00440-5}

\bibitem[Rasch et~al., 2007]{rasch2007odor}
Rasch, B., B{\"u}chel, C., Gais, S., \& Born, J. (2007).
\newblock Odor cues during slow-wave sleep prompt declarative memory
  consolidation.
\newblock {\em Science}, 315(5817), 1426--1429.
\newblock \url{https://doi.org/10.1126/science.1138581}

\bibitem[Rosenberg \& Van~Hout, 2013]{rosenberg2013american}
Rosenberg, R.~S. \& Van~Hout, S. (2013).
\newblock The american academy of sleep medicine inter-scorer reliability
  program: sleep stage scoring.
\newblock {\em Journal of clinical sleep medicine}, 9(1), 81--87.
\newblock \url{https://doi.org/10.5664/jcsm.2350}

\bibitem[Rudoy et~al., 2009]{rudoy2009strengthening}
Rudoy, J.~D., Voss, J.~L., Westerberg, C.~E., \& Paller, K.~A. (2009).
\newblock Strengthening individual memories by reactivating them during sleep.
\newblock {\em Science}, 326(5956), 1079--1079.
\newblock \url{https://doi.org/10.1126/science.1179013}

\bibitem[Supratak et~al., 2017]{supratak2017deepsleepnet}
Supratak, A., Dong, H., Wu, C., \& Guo, Y. (2017).
\newblock Deepsleepnet: A model for automatic sleep stage scoring based on raw
  single-channel eeg.
\newblock {\em IEEE Transactions on Neural Systems and Rehabilitation
  Engineering}, 25(11), 1998--2008.
\newblock \url{https://doi.org/10.1109/TNSRE.2017.2721116}

\bibitem[Supratak \& Guo, 2020]{supratak2020tinysleepnet}
Supratak, A. \& Guo, Y. (2020).
\newblock Tinysleepnet: An efficient deep learning model for sleep stage
  scoring based on raw single-channel eeg.
\newblock {\em 2020 42nd Annual International Conference of the IEEE
  Engineering in Medicine \& Biology Society (EMBC)}, 641--644.
\newblock \url{https://doi.org/10.1109/EMBC44109.2020.9176741}

\bibitem[Vallat \& Walker, 2021]{vallat2021open}
Vallat, R. \& Walker, M.~P. (2021).
\newblock An open-source, high-performance tool for automated sleep staging.
\newblock {\em Elife}, 10, e70092.
\newblock \url{https://doi.org/https://doi.org/10.7554/eLife.70092}

\end{thebibliography}






\end{document}